\documentclass[%
reprint,
amsmath,amssymb,
aps,longbibliography,
]{revtex4-2}

\usepackage{xcolor}
\usepackage{url}
\usepackage[colorlinks=true,linkcolor=blue,urlcolor=blue,anchorcolor=blue,citecolor=blue,bookmarksnumbered]{hyperref}
\usepackage{graphicx}
\usepackage{dcolumn}
\usepackage{bm}

\begin{document}

\title{Non-Abelian holonomic transformations in digitally coupled acoustic waveguides guided by the global adiabatic criterion}
\author{Jin-Kang~Guo$^{1}$}\author{Jia~Li$^{2}$}
\author{Jin-Lei~Wu$^{2}$}\email[]{jlwu517@zzu.edu.cn}
\author{Chuan-Cun~Shu$^{1}$}\email[]{cc.shu@csu.edu.cn}
\affiliation{$^{1}$School of Physics, Central South University, Changsha 410083, China}
\affiliation{$^{2}$Quantum Information Institute, School of Physics, Zhengzhou University, Zhengzhou 450001, China}


\begin{abstract}
An acoustic platform is validated for implementing compact non-Abelian holonomic transformations (NHTs) guided by a global adiabatic criterion~(GAC). A tripod model is mapped onto a digitally coupled four-waveguide structure, where designed coupling envelopes and an acoustically-induced-transparency phase-control module implement a two-stage phase-stitched holonomic evolution. Compared with a reference Gaussian envelope, the GAC-guided power-law profile flattens the spatial distribution of the global nonadiabatic burden, thereby providing a quantitative basis for compact acoustic implementation. Full-wave simulations show Pauli-$X$ and Hadamard-type target transformations, with excellent agreement between the extracted normalized intensities and analytical coupled-mode predictions. These target responses are obtained with half the coupling length required by the reference Gaussian implementations. More uniquely, the same phase-stitched structure also supports unidirectional acoustic mode conversion, which is closely related to a reduced two-mode non-Hermitian picture associated with an encircled exceptional point~(EP). These results validate acoustic NHTs as a robust geometric route for compact wave control, establish the GAC as a powerful guideline for fast adiabatic transport in digitally coupled systems, and further demonstrate that the same phase-stitched architecture supports unidirectional mode conversion through EP-assisted branch selection.
\end{abstract}

\maketitle

\section{Introduction}

Geometric control provides an appealing route to logic operations because the target transformation is determined by the global evolution path rather than by local dynamical details~\cite{Berry1984BerryPhase,Simon1983Holonomy,JZhang2023PR,YLiang2023SCIS,ZLShan2025SB}. Related coherent-control strategies have also been extensively explored in molecular, atomic, and hybrid light--matter systems, including stimulated Raman adiabatic passage (STIRAP), rotational-state engineering, pulsed Raman control, and molecular-polariton manipulation~\cite{Bergmann1998STIRAP,Vitanov2017STIRAP,Shu2009MolecularSTIRAP,Hong2025MolecularRotationControl,Jian2025PulsedJumpRaman,Fan2023MolecularPolaritonControl}. In particular, non-Abelian holonomic transformations (NHTs) supported by degenerate dark-state subspaces enable noncommuting transformations in the logical subspace and have become a central theme in coherent control~\cite{Wilczek1984NonAbelian,Zanardi1999Holonomic,Duan2001Holonomic,Leibfried2003Nature,Sjoqvist2012NHQC,Xu2018SingleLoopNHQC,Sun2022NP,Zhang2022NatPhoton,JXie2026NC}. Owing to the formal equivalence between Schr\"odinger-type evolution and coupled-mode transport~\cite{Enrich2016ROPP,ZXChen2025SB}, these non-Abelian holonomic formalisms can be transferred to classical-wave platforms. Acoustic waveguides provide a macroscopic setting in which coherent transport, adiabatic following, and geometric transformations can be engineered and directly visualized~\cite{Chen2022AcousticBraiding,Barlas2020TopologicalBraiding,Shen2019AcousticLAP,Tang2022AcousticSTAP,Tang2023AcousticBeamSplit,Wu2022AcousticNHQT,Liu2022AcousticGeometricPhase}.

Existing acoustic implementations inspired by the adiabatic passage rely on the local adiabatic condition, which requires pointwise slow modulations of the coupling profiles along the propagation direction~\cite{LYang2019JASA,Chen2021AcousticLandauZener,Chen2022AcousticPump,ZGuan2024PRApplied,ZCheng2025NC,QMo2025PRL,chen2026topologicalphononics}. This requirement becomes restrictive in compact phase-engineered devices. For holonomic transformations such as the Pauli-$X$ and Hadamard operations, the acoustic structure must combine controlled coupling evolution with a stitched $\pi$ phase modulation between two coupling stages. Maintaining dark-state following under this combined length and phase constraint is therefore a nontrivial design problem.
Several strategies have been developed to accelerate adiabatic transport, including shortcut engineering, structure reforming, and coupling tailoring~\cite{Wu2017ShortcutAdiabatic,Liu2017SuperadiabaticHQC,Vasilev2009OptimumSTIRAP,Tang2022AcousticSTAP,Tang2022AcousticRFSTIRAP,Yao2024AcousticFastSSH}.
Despite improving local transfer efficiency, these methods are typically confined to pointwise optimizations that introduce extra experimental overhead and fragility against parameter drifts, while leaving the global accumulation of nonadiabatic effects entirely unquantified. Recent work has proposed the global adiabatic criterion (GAC) for guiding accelerated adiabatic evolution~\cite{Xiao2025GAC,wu2026globaladiabaticcriterionfast,guo2026nonabelianthoulesspumpingbased}, where the criterion quantifies, in a global and integrated manner, the total nonadiabatic leakage and its nonuniform distribution, thus enabling the systematic optimization of driving protocols rather than fragmentary local adjustments. This perspective naturally motivates its application as a timing and envelope-design principle for compact acoustic holonomic transformations.

\begin{figure*}[htp]
	\centering
	\includegraphics[width=0.66\linewidth]{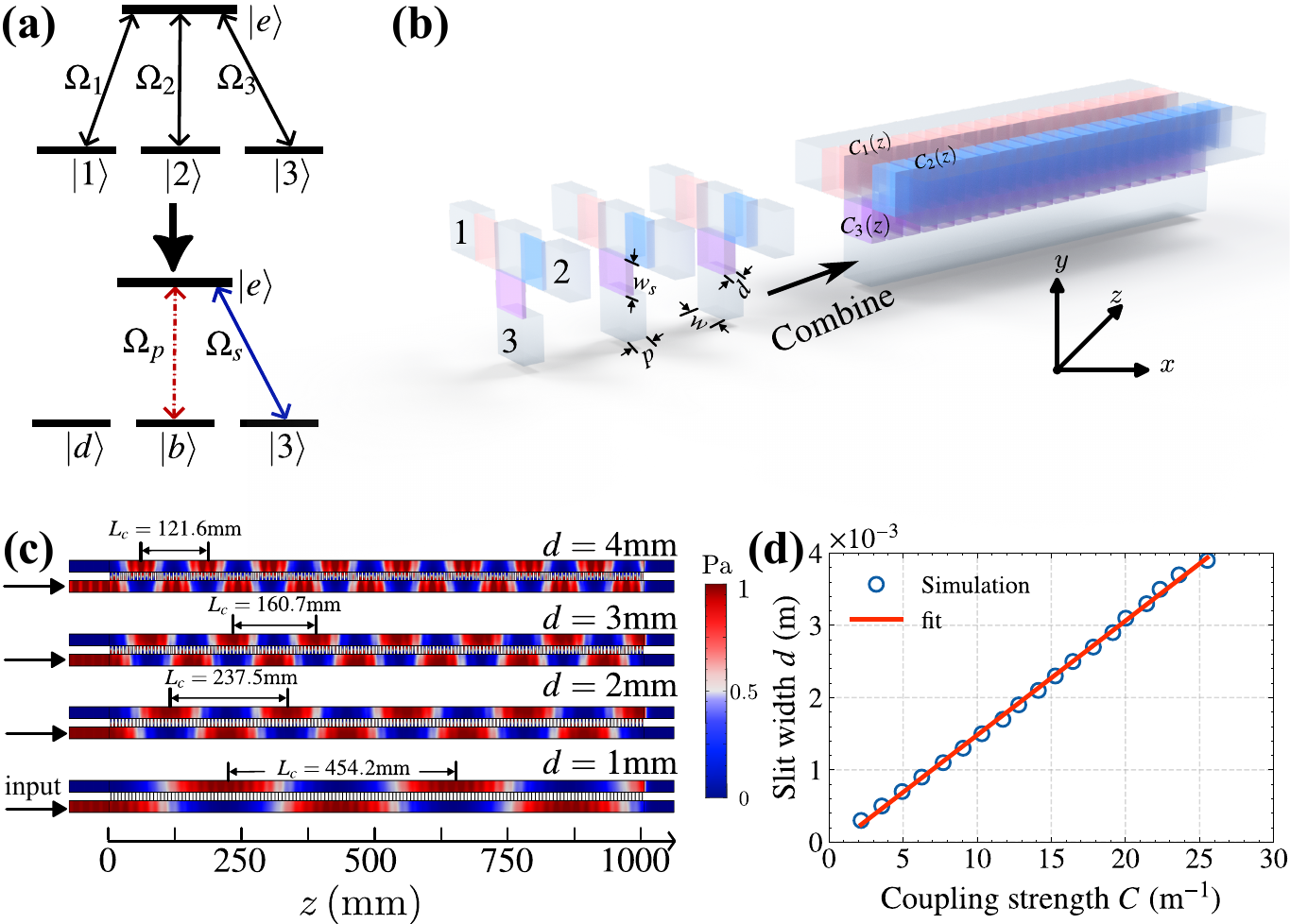}
	\caption{(a) Schematic of the four-level system and its bright--dark reduction, where $|d\rangle$ is decoupled while $|b\rangle$ and $|3\rangle$ are coupled to $|e\rangle$. (b) Acoustic implementation in a slit-coupled four-waveguide structure, where the spatially varying couplings $C_1(z)$, $C_2(z)$, and $C_3(z)$ are combined into a continuous device. (c) Simulated field distributions for different slit widths $d$, showing the dependence of the effective coupling. (d) Calibration curve of slit width $d$ versus coupling strength $C$, where the simulation data are well described by a linear fit.}
	\label{fig:Fig1}
\end{figure*}
In this work, we present an analytical and full-wave study of acoustic NHTs guided by the GAC. A slit-induced digitally-coupled four-waveguide structure is used to emulate a tripod model with a twofold-degenerate dark subspace, and a resonant phase-control module supplies the required phase stitching~\cite{Cheng2019AIT,Porter2022AITATS}. We compare a conventional Gaussian adiabatic-passage envelope~\cite{Shen2019AcousticLAP} with our GAC-guided power-law envelope under the same slit-coupling calibration.
This comparison demonstrates a $50\%$ reduction of the active coupling length enabled by our GAC design. We further show that the Hadamard-type NHT configuration incorporating with an imposed loss contrast supports a directional mode conversion with suppressed reverse transmission, connecting the acoustic holonomic platform to a functional unidirectional transport control~\cite{Liang2009AcousticDiode,Liang2010AcousticRectifier,Shen2016AsymmetricTransmission,Liu2018PTSymmetry,Liu2020SingleSidedBeamSplit,Song2019HighIndexPrism,Tang2023AcousticDoubleFSTIRAP}. This unidirectional response is closely associated with an encircled exceptional point~(EP) in a reduced non-Hermitian branch-selection picture~\cite{Doppler2016EPEncircling,XLZhang2018PRX,Li2020HamiltonianHopping,Shu2022FastEncirclement,Guo2023ExceptionalNonAbelian,ALi2023NNano,Arkhipov2024RestoringAdiabatic,ZLShan2024PRL,ZLi2026NP,PXue2026PRL,MRYun2026PRA}, where the coupling-ratio path and the loss contrast provide the mechanism for directional conversion and attenuation. This work therefore validates the GAC as an efficient and broadly applicable design principle for accelerating adiabatic evolution in classical wave systems, and demonstrates that acoustic waveguides offer a scalable setting for emulating non-Abelian geometric gates and extending the additional functionality of unidirectional control enabled by the same phase-stitched configuration.

\section{Theoretical Framework}
\subsection{Four-level description}
We begin with a four-level tripod model in the context of an atomic system shown in Fig.~\ref{fig:Fig1}(a),  where the three lower states $|1\rangle$, $|2\rangle$, and $|3\rangle$ are coupled to an auxiliary upper state $|e\rangle$ with time-dependent transition Rabi frequencies $\Omega_{1,2,3}(t)$, respectively. The Hamiltonian reads
\begin{equation}
    \hat{H}(t)=\sum_{j=1}^{3}\Omega_j(t)\,|e\rangle\langle j|+\mathrm{H.c.}.
\end{equation}
It is convenient to parameterize the two couplings associated with $|1\rangle$ and $|2\rangle$ to
$\Omega_1(t)=\Omega_p(t)\sin\varphi$ and $\Omega_2(t)=\Omega_p(t)\cos\varphi$ with the constant mixed angle $\varphi$ setting the target transformation, and to introduce the bright and dark superpositions in the subspace $\{|1\rangle,|2\rangle\}$ as $|b\rangle=\sin\varphi\,|1\rangle+\cos\varphi\,|2\rangle$ and $|d\rangle=\cos\varphi\,|1\rangle-\sin\varphi\,|2\rangle$. The dark state $|d\rangle$ remains decoupled from the system, while the remaining $|b\rangle$ and $|3\rangle$ are coupled to $|e\rangle$ with couplings $\Omega_p(t)$ and $\Omega_s(t)\equiv\Omega_3(t)$, respectively. Rewriting these couplings as $\Omega_p(t)=\Omega(t)\sin\theta(t)$ and $\Omega_s(t)=\Omega(t)\cos\theta(t)e^{-i\phi_s(t)}$, we obtain another instantaneous dark state given by $|\psi_0(t)\rangle=\cos\theta(t)|b\rangle-\sin\theta(t)e^{i\phi_s(t)}|3\rangle$. The evolution is thus confined to a twofold-degenerate dark manifold spanned by $\{|d\rangle,|\psi_0(t)\rangle\}$.

For the cyclic protocol considered here, the phase of $|3\rangle\leftrightarrow|e\rangle$ coupling channel is taken as $\phi_s(t)=0$ in the first half-cycle but $\phi_s(t)=\gamma$ in the second half-cycle, so that $\gamma$ denotes the relative phase stitching between the two stages. Under such a cyclic evolution, the system is driven from $|b\rangle$ to $-|3\rangle$ in the first half-cycle and then returned to the dark manifold in the second half-cycle. As a result, the evolution in the computational subspace $\{|b\rangle,|d\rangle\}$ is purely geometric~\cite{Duan2001Holonomic,Liu2017SuperadiabaticHQC}. The state $|d\rangle$ remains unchanged, while $|b\rangle$ acquires a holonomic phase determined by $\gamma$. The resulting unitary propagator in the basis $\{|b\rangle,|d\rangle\}$ can therefore be written as (up to a global phase)
\begin{equation}
    U=
    \begin{pmatrix}
        e^{i\gamma} & 0\\
        0 & 1
    \end{pmatrix}.
\end{equation}
 Different Pauli-$X$ and Hadamard-type transformations in the logical basis $\{|1\rangle,|2\rangle\}$ are then obtained by selecting the mixing angle $\varphi$.

\subsection{Implementation in acoustic waveguides}
The corresponding acoustic implementation uses the slit-induced digitally-coupled four-waveguide structure shown in Fig.~\ref{fig:Fig1}(b), where the acoustic pressure amplitudes play the role of state amplitudes. They obey a coupled-mode theory (CMT) description along the propagation direction $z$, with slit-induced coupling strengths $C_j(z)$ corresponding to the designed coupling envelopes~\cite{Wu2022AcousticNHQT,Yao2024AcousticFastSSH,Tang2023AcousticDoubleFSTIRAP,Tang2023AcousticBeamSplit,Tang2022AcousticRFSTIRAP,Tang2022AcousticSTAP,Shen2019AcousticLAP}. Physically, different from continuous interconnection among acoustic cavities~\cite{ZGuan2024PRApplied,Chen2021AcousticLandauZener,ZCheng2025NC,QMo2025PRL,Chen2022AcousticPump}, the periodically arranged slits provide aperture-mediated coherent modal coupling between neighboring acoustic waveguides, leading to digitally controlled inter-channel energy exchange with a coupling length determined by the slit width, rather than a local-resonator-type response.
After the same bright-dark reduction in the $\{|1\rangle,|2\rangle\}$ subspace, the acoustic system holds an effective spatially modulated Hamiltonian in the basis $\{|b\rangle,|3\rangle,|e\rangle\}$
\begin{equation}
    \hat{H}_{\mathrm{eff}}(z)=
    \begin{pmatrix}
        0 & 0 & C_p(z) \\
        0 & 0 & C_s(z) \\
        C_p^*(z) & C_s^*(z) & 0
    \end{pmatrix},
\end{equation}
where $C_s(z)\equiv C_3(z)$. Its instantaneous dark state is
\begin{equation}
    |\psi_0(z)\rangle=\cos\theta(z)|b\rangle-\sin\theta(z)e^{i\phi_s(z)}|3\rangle,
\end{equation}
which underlies the phase-stitched holonomic evolution studied below.

The full acoustic structure is implemented with a three-dimensional slit-coupled four-waveguide device, and all full-wave simulations are carried out based on the finite-element method.
The surrounding solid structure is treated as acoustically rigid, so that the waveguide walls and slit boundaries are modeled by hard-wall boundary conditions. In the full device, the waveguide width, inter-waveguide wall thickness, and slit period are fixed at $w=10$~mm, $w_s=8$~mm, and $p=7$~mm, respectively, and the target couplings are encoded by the slit-width profile $d(z)$. To establish the geometric mapping from slit width to effective coupling, we calibrate an auxiliary two-waveguide unit for different discrete slit widths and extract the corresponding coupling strengths from the coupling lengths, as shown in Fig.~\ref{fig:Fig1}(c). The resulting calibration curve in Fig.~\ref{fig:Fig1}(d) is well fitted in the working range by the linear relation $d(C)=aC+b$, with $a=1.5847\times10^{-4}~{\rm m}^{2}$ and $b=-9.6245\times10^{-5}~{\rm m}$. In the present design, the slit width is limited to $d_{\max}=4$~mm, corresponding to a maximum usable coupling strength of $C_{\max}=25.85$~m$^{-1}$. This fitted relation is then used to convert the designed coupling envelopes into the slit-width distribution of the full four-waveguide device.

\begin{figure}[b]
	\centering
	\includegraphics[width=0.95\linewidth]{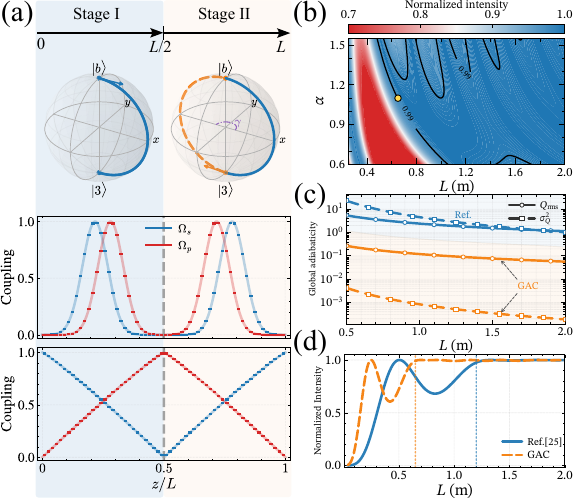}
	\caption{GAC-guided envelope design and normalized target-output intensity.
		(a) Two-stage phase-stitched protocol and the reference Gaussian and GAC-guided coupling profiles, including their slit-period-sampled realizations.
		(b) Normalized target-output intensity of the GAC-guided power-law profiles in the $(\alpha,L)$ parameter space. The color scale denotes the normalized intensity, the black contour marks $0.99$, and the marker denotes the selected working point.
		(c) Merged global adiabaticity metrics $Q_{\rm rms}$ and $\sigma_Q^2$ for the reference Gaussian and GAC-guided profiles.
		(d) Normalized target-output intensity versus the total active coupling length $L$, comparing the reference Gaussian~\cite{Shen2019AcousticLAP} and GAC-guided profiles.}
	\label{fig:Fig2}
\end{figure}
\subsection{Nonadiabaticity factor}
To assess the validity of dark-state following, we consider the instantaneous adiabatic condition in the general form
$|\langle\phi_m(z)|\partial_z|\psi_0(z)\rangle|/\Delta E_m(z)\ll 1$,
where $|\phi_m(z)\rangle$ $(m\neq0)$ are bright instantaneous eigenstates and
$\Delta E_m(z)=|E_m(z)-E_0(z)|$ are the corresponding eigenvalue gaps.
For the effective $\Lambda$-type subsystem considered here, we obtain the nonadiabaticity factor~\cite{Xiao2025GAC,wu2026globaladiabaticcriterionfast,guo2026nonabelianthoulesspumpingbased}
\begin{equation}
    Q(z)\equiv\sum_m\frac{|\langle\phi_m(z)|\partial_z|\psi_0(z)\rangle|}{\Delta E_m(z)}=\frac{|\partial_z\theta(z)|}{\sqrt{2}\,C(z)},
\end{equation}
where $C(z)=\sqrt{|C_p(z)|^2+|C_s(z)|^2}$.
For a propagation interval of length $\ell$, we define the mean nonadiabaticity as
\begin{equation}
    \overline{Q}=\frac{1}{\ell}\int_0^{\ell}Q(z)\,dz,
\end{equation}
and the nonadiabaticity variance as
\begin{equation}
    \sigma_Q^2=\frac{1}{\ell}\int_0^{\ell}\left[Q(z)-\overline{Q}\right]^2dz,
\end{equation}
The corresponding root-mean-square is $Q_{\rm rms}=\sqrt{\overline{Q}^{\,2}+\sigma_Q^2}$.
Here, $Q_{\rm rms}$ characterizes the overall global nonadiabatic burden, whereas $\sigma_Q^2$ measures its spatial inhomogeneity. In our two-stage protocol of acoustic NHTs, these quantities are evaluated over a single coupling stage and are used as the global adiabaticity metrics for comparing different coupling schemes.

\section{GAC-guided protocol design}
\subsection{Holonomic evolution protocol}
In the present work, the cyclic holonomic evolution is implemented by two coupling stages connected through a phase-control module, as illustrated in Fig.~\ref{fig:Fig2}(a). In the envelope comparison below, $L$ denotes the total active coupling length of the two coupling stages, and each stage has length $L_h=L/2$. We use $\tau\in[0,L_h]$ as the local propagation coordinate within a stage. The stitched phase on the $|3\rangle\leftrightarrow|e\rangle$ coupling channel is taken as
\[
\phi_s =
\begin{cases}
0, & \text{Stage I},\\
\gamma, & \text{Stage II},
\end{cases}
\]
with $\gamma=\pi$ in the present design. The mixing angle $\varphi$ fixes the target holonomic transformation in the logical basis; specifically, $\varphi=\pi/4$ and $\varphi=\pi/8$ correspond to the Pauli-$X$ and Hadamard transformations, respectively. The theoretically designed coupling envelopes are smooth functions of $\tau$. In the practical discrete-slit coupled acoustic structure, these envelopes are implemented by sampling the local coupling strength once per period $p$, producing the digital profiles shown in Fig.~\ref{fig:Fig2}(a). This sampling step maps the target envelopes onto slit widths while keeping the same ordering and relative overlap of the couplings.

To benchmark the role of GAC, we compare the GAC-guided envelope with a reference Gaussian profile used for a typical acoustic adiabatic-passage experiment~\cite{Shen2019AcousticLAP}. In the normalized stage coordinate $u=\tau/L_h$, the reference profile is written as
$C_s=C_0\exp[-(u-u_s)^2/\sigma_u^2]$ and $C_p=C_0\exp[-(u-u_p)^2/\sigma_u^2]$.
Involved quantities are defined by $u_s=(z_0-s)/(2z_0)$, $u_p=(z_0+s)/(2z_0)$, and $\sigma_u=\sigma/(2z_0)$, with parameters $z_0=430~\mathrm{mm}$, $2s=110~\mathrm{mm}$, and $\sigma=127~\mathrm{mm}$~\cite{Shen2019AcousticLAP}. The GAC-guided scheme instead uses the power-law pair $C_p=C_0[1-(1-u)^\alpha]$ and $C_s=C_0(1-u^\alpha)$. Following the GAC-based selection strategy of Ref.~\cite{Xiao2025GAC}, a scan of the power-law family identifies $\alpha=1.10$ as the working exponent for the present acoustic parameters. Here $C_0=C_{\max}$ is the peak coupling strength used for both schemes. Although the two protocols use the same two-step structure, they differ substantially in the spatial allocation of coupling strength over each step~[see the two lower panels of Fig.~\ref{fig:Fig2}(a)]. The slit-period sampling discretizes the designed envelopes while preserving their global ordering in the acoustic structure.

\begin{figure}[b]
	\centering
	\includegraphics[width=0.95\linewidth]{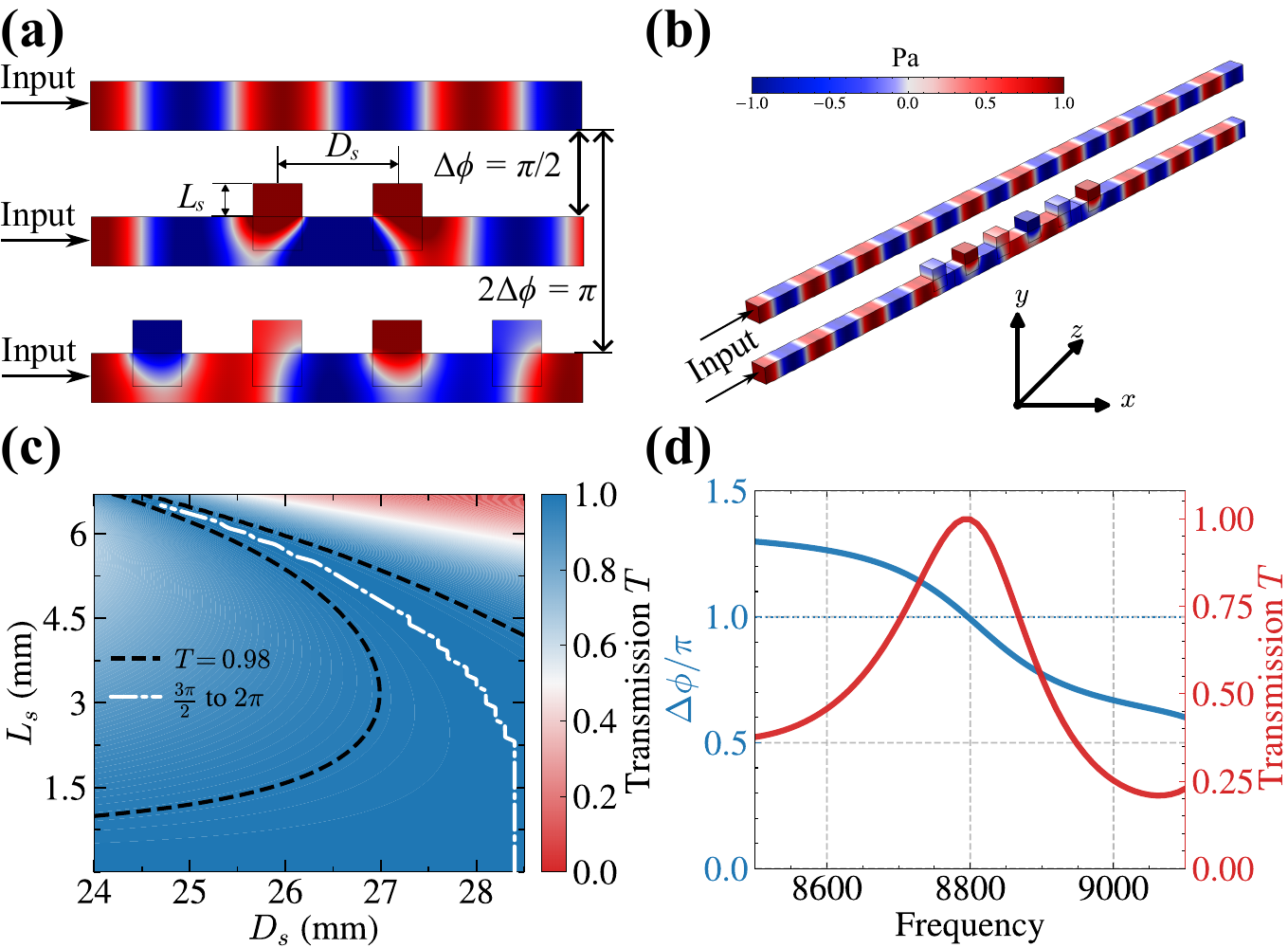}
	\caption{(a) Schematic of the phase-shifting unit formed by two side-coupled resonators, with $D_s$ and $L_s$ denoting the key geometric parameters. (b) Three-dimensional acoustic realization of the phase-shifting structure. (c) Simulated transmission map in the $(D_s,L_s)$ parameter space, where the dashed contour marks the phase-shift range from $3\pi/2$ to $2\pi$ and the high-transmission contour marks $T=0.98$. (d) Simulated phase shift $\Delta\phi/\pi$ and transmission $T$ versus frequency for the chosen design.}
	\label{fig:Fig3}
\end{figure}
\begin{figure*}[t]
	\centering
	\includegraphics[width=0.88\textwidth]{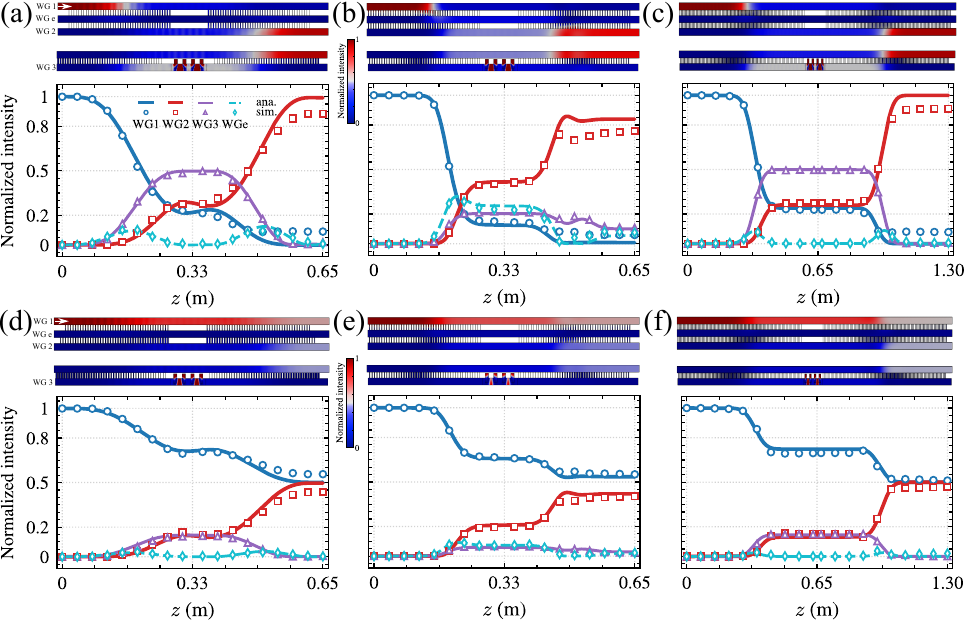}
	\caption{Full-wave validation of the acoustic Pauli-$X$ and Hadamard transformations. (a--c) Pauli-$X$ transformation for the GAC-guided profile with an active coupling length of $65\mathrm{cm}$, the Gaussian profile with the same active coupling length of $65\mathrm{cm}$, and the longer Gaussian reference with an active coupling length of $130~\mathrm{cm}$, respectively. (d--f) Corresponding Hadamard transformation for the same three implementations. In each panel, the upper map shows the full-wave normalized acoustic intensity distribution, and the lower plot shows the four-waveguide normalized intensities as functions of the physical distance $z$. Solid curves denote the analytical coupled-mode predictions, while symbols denote the full-wave simulation results.}
	\label{fig:Fig4}
\end{figure*}
We further examine the normalized output intensity in the target waveguide and the global adiabaticity metrics for the continuous envelopes as functions of the total active coupling length $L$. Figure~\ref{fig:Fig2}(b) maps the normalized target-output intensity of the GAC-guided power-law profiles in the $(\alpha,L)$ parameter space. The color scale denotes the normalized intensity, the black contour marks the $0.99$ level, and the circle marker denotes the selected working point $(\alpha,L)=(1.10,65\mathrm{cm})$. Figure~\ref{fig:Fig2}(c) presents the merged global adiabaticity metrics $Q_{\rm rms}$ and $\sigma_Q^2$ for the reference Gaussian and GAC-guided profiles. These metrics are evaluated over one coupling stage, $\ell=L_h=L/2$, and plotted against the total active coupling length $L$. Compared with the reference Gaussian profile~\cite{Shen2019AcousticLAP}, the GAC-guided profile gives a far smaller $Q_{\rm rms}$ and a greatly reduced $\sigma_Q^2$ over the considered length range, indicating a much lower global nonadiabatic burden and a more uniform spatial distribution of the adiabatic condition. The corresponding normalized target-output intensity comparison in Fig.~\ref{fig:Fig2}(d) shows that the normalized target-output intensity increases with $L$ for both profiles, while the GAC-guided profile reaches the target-output level at a substantially shorter total active coupling length than the reference Gaussian profile~\cite{Shen2019AcousticLAP}. These results show that the GAC provides a practical design guideline for reducing the device length of acoustic NHTs.

\subsection{Phase shift protocol}
To implement the required phase control in a realistic acoustic structure, we design a phase-shifting unit by employing a side-branch-resonator module that operates in the transparency regime rather than introducing a direct geometric discontinuity. Such transparency-assisted transport has been discussed in acoustic resonant-channel systems in terms of acoustically induced transparency and its crossover to Autler-Townes-type splitting~\cite{Cheng2019AIT,Porter2022AITATS}. In the present design, based on the operating principle summarized in Fig.~\ref{fig:Fig3}(a), the basic configuration of the phase-shifting unit is shown in Fig.~\ref{fig:Fig3}(b). A single module provides an approximate phase delay of $\Delta\phi=\pi/2$, while two cascaded modules yield the total phase shift required by the protocol. The joint dependence of transmission and phase delay on the structural parameters $L_s$ and $D_s$ is presented in Fig.~\ref{fig:Fig3}(c), where an overlap region is identified. The target phase response and high transmission are simultaneously satisfied. The working geometry is selected as $L_s=6.67~\mathrm{mm}$ and $D_s=24.4~\mathrm{mm}$. Here $L_s$ mainly tunes the resonant path length and hence the phase delay, while $D_s$ adjusts the coupling and interference condition of the side branches and thereby affects the transparency window. Figure~\ref{fig:Fig3}(d) shows the frequency sweep for the chosen design. At the selected operating acoustic-wave frequency $f=8800~\mathrm{Hz}$, the two cascaded modules provide a total phase delay of $\pi$, while the selected point remains on the high-transmission contour $T_{8800}=0.99$.

\section{Full-Wave Demonstration}

\subsection{Non-Abelian holonomic transformations}
With the calibrated phase-control module incorporated into the four-waveguide structure, we perform full-wave simulations of the complete implementations at $f=8800~\mathrm{Hz}$. Figure~\ref{fig:Fig4} compares six cases under the same operating frequency, waveguide cross section, slit-coupling calibration, and stitched phase $\gamma=\pi$. Panels (a)--(c) show the Pauli-$X$ transformation with $\varphi=\pi/4$, and panels (d)--(f) show the Hadamard transformation with $\varphi=\pi/8$. For each transformation, the three columns correspond to the GAC-guided profile with $L=65~\mathrm{cm}$, the Gaussian profile compressed to the same active coupling length $L=65~\mathrm{cm}$, and the longer Gaussian reference with $L=130~\mathrm{cm}$. The quoted active coupling length refers to the two coupling stages and excludes the phase-control module. The input is launched from the wavegudie 1 (WG1) as the fundamental port mode. In each panel, the upper map shows the full-wave normalized acoustic intensity distribution, and the lower plot compares the four-waveguide normalized intensities with the analytical coupled-mode predictions. The normalized intensity in WG$j$~($j=1,2,3,e$) is defined as $I_j(z)=\int_{S_j(z)}|p(\mathbf r)|^2\,dS\,\big/\,\sum_k\int_{S_k(z)}|p(\mathbf r)|^2\,dS$, where $S_j(z)$ denotes the transverse cross section of WG$j$, and the sum runs over $k=\{\mathrm{WG}j\}$.

The two target transformations exhibit the same length-dependent contrast between the GAC-guided and Gaussian profiles. For the Pauli-$X$ case, Fig.~\ref{fig:Fig4}(a) shows that the GAC-guided implementation transfers most of the input intensity from WG1 to the target channel WG2 within $65~\mathrm{cm}$. The compressed Gaussian implementation in Fig.~\ref{fig:Fig4}(b) has not yet settled into the target-output distribution at the same length and retains stronger residual population in the non-target channels. This behavior is consistent with Fig.~\ref{fig:Fig2}(d), where the Gaussian profile at $L=65~\mathrm{cm}$ still lies in a strongly length-dependent nonadiabatic region, while the selected GAC-guided working point has reached the high target-output region. Increasing the Gaussian active coupling length to $130~\mathrm{cm}$ achieves the target Pauli-$X$ output in Fig.~\ref{fig:Fig4}(c). The same comparison holds for the Hadamard case in Figs.~\ref{fig:Fig4}(d)--\ref{fig:Fig4}(f), where the GAC-guided profile produces the intended modal-intensity splitting within $65~\mathrm{cm}$, the compressed Gaussian profile remains more affected by nonadiabatic leakage, and the longer Gaussian reference recovers the expected Hadamard-type output. The full-wave extracted intensities follow the coupled-mode predictions in all six cases, showing that the observed difference arises from the spatial allocation of the coupling profiles rather than from the extraction procedure. Together with the global adiabaticity metrics in Fig.~\ref{fig:Fig2}(c), these results show that the GAC-guided envelope reaches the target-output region at a shorter active coupling length, whereas the compressed Gaussian profile~\cite{Shen2019AcousticLAP} remains limited by stronger nonadiabatic leakage at the same length.

\begin{figure}[t]
    \centering
    \includegraphics[width=\linewidth]{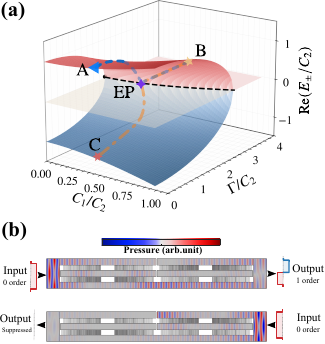}
    \caption{Unidirectional acoustic mode conversion enabled by the phase-stitched holonomic acoustic structure. (a) Reduced two-mode non-Hermitian Riemann-surface/EP schematic. (b) Full-wave directional mode-conversion pressure fields. Under forward excitation, the incident zero-order mode is converted into the first-order output mode, whereas under reverse excitation the transmitted output is strongly suppressed.}
    \label{fig:Fig5}
\end{figure}
\subsection{Unidirectional mode conversion}
Figure~\ref{fig:Fig5} examines a unidirectional mode-conversion response obtained from the Hadamard-type phase-stitched four-waveguide structure. The simulation is performed at $f=8800~\mathrm{Hz}$ with the same waveguide cross section and slit-coupling calibration used above, while the coupling layout and added loss are selected so that the reduced parameters connect to the EP-related branch structure in Fig.~\ref{fig:Fig5}(a). The full-wave pressure fields in Fig.~\ref{fig:Fig5}(b) show two opposite excitation directions. Under forward excitation, an incident zero-order transverse mode (symmetric superposition $|1\rangle+|2\rangle$) is converted into a first-order output mode ($|1\rangle-|2\rangle$) as indicated by the anti-symmetric pressure profile at the output. Under reverse excitation, however, the transmitted field is strongly suppressed, giving a unidirectional mode-conversion response.

The Riemann surface of a non-Hermitian system in Fig.~\ref{fig:Fig5}(a) provides a reduced description of the branch selection associated with this phase-stitched four-waveguide configuration. The ratio $C_1/C_2$ is inherited from the original four-level coupling structure and fixes the bright--dark mixing in the $|1\rangle,|2\rangle$ subspace, with $C_1/C_2=\tan\varphi$ under the parameterization $C_1=C_p\sin\varphi$ and $C_2=C_p\cos\varphi$. The parameter \(\Gamma\) represents the loss applied to WG1 and WGe at the junction between the two concatenated Hadamard sections. In this reduced parameter space, the normalized eigenvalue branches are written as $E_\pm/C_2=\pm[1+(C_1/C_2)^2-(\Gamma/2C_2)^2]^{1/2}$, whose real parts form the two-sheet Riemann surface in the $(C_1/C_2,\Gamma/C_2)$ plane~\cite{Doppler2016EPEncircling,Li2020HamiltonianHopping,Shu2022FastEncirclement,Guo2023ExceptionalNonAbelian,Arkhipov2024RestoringAdiabatic}.  The EP curve is given by $\Gamma/C_2=2\sqrt{1+(C_1/C_2)^2}$, and the marked point indicates a representative coalescence point on this curve. The $A\rightarrow B\rightarrow C$ trajectory connects the state associated with the zero-order input mode on the upper sheet to that associated with the first-order output mode on the lower sheet. Near the large-loss boundary at $B$, a rapid parameter hop between convergent boundary eigenstates connects the two sheets, while the attenuation contrast introduced by $\Gamma$ makes the branch toward $C$ dominant. This loss-assisted branch selection is analogous to Hamiltonian-hopping-assisted EP dynamics, where boundary hopping and loss contrast can suppress path-dependent dissipation~\cite{Li2020HamiltonianHopping,Shu2022FastEncirclement}. The reduced picture therefore links the imposed loss contrast and the coupling-ratio path to the full-wave unidirectional conversion observed in Fig.~\ref{fig:Fig5}(b).

\section{Conclusion}
In summary, we have proposed and numerically validated a macroscopic acoustic platform for non-Abelian holonomic transformations guided by the GAC. Starting from a four-level model, we mapped the target evolution onto a slit-induced digitally coupled four-waveguide structure and calibrated the relation between the geometric slit width and the effective coupling strength. The resulting device supports a twofold-degenerate dark subspace, and the Pauli-$X$ and Hadamard transformations are implemented through a two-stage propagation protocol with phase stitching on the third coupling channel. The GAC-guided power-law profile gives smaller $Q_{\rm rms}$ and $\sigma_Q^2$ than the reference Gaussian profile, corresponding to a lower global nonadiabatic burden and a more uniform spatial distribution of the adiabatic condition. For both the Pauli-$X$ and Hadamard transformations, the GAC-guided devices realize the target transformations with just a half of that required for the reference Gaussian implementation~\cite{Shen2019AcousticLAP} to reach comparable output behaviors. Furthermore, we show that the same phase-stitched acoustic structure also supports unidirectional mode conversion, which is exactly interpreted through a reduced two-mode non-Hermitian picture with an EP. These results therefore link the GAC design with a compact acoustic implementation of non-Abelian holonomic transformations, establishing a unified framework that integrates global adiabatic engineering, digitally coupled waveguide networks, and non-Hermitian branch-selection physics, thereby offering a versatile and scalable route to both geometric quantum simulation and unidirectional wave control in classical platforms.

\begin{acknowledgments}
	The authors acknowledge financial support from the  National Natural Science Foundations of China (NSFC) under Grants No. 62571494, No. 12304407, and No. 12274470, from the
	China Postdoctoral Science Foundation under Grants No. 2023TQ0310 and No. GZC20232446, and from the Natural Science Foundation of Henan Province under Grant No. 262300421244.
\end{acknowledgments}

\bibliography{ref}% Produces the bibliography via BibTeX.

\end{document}